\documentstyle[preprint,aps]{revtex}
\newcommand{\clt}        {$\rm^{1}$}
\newcommand{\gsi}        {$\rm^{2}$}

\begin{document}

\title{Investigation of background subtraction techniques 
for high mass dilepton physics}

\date{\today}

\author{P.~Crochet\clt~and P.~Braun-Munzinger\gsi}

\address{
\clt~Laboratoire de Physique Corpusculaire, IN2P3/CNRS
and Universit\'{e} Blaise Pascal, Clermont-Ferrand, France\\
\gsi~Gesellschaft f\"ur Schwerionenforschung, Darmstadt, Germany
}

\maketitle

\begin{abstract}
The signals in high mass dilepton spectroscopy with nucleus-nucleus 
collisions at collider energies are superimposed on a generally large 
combinatorial background. 
Because this background contains a significant correlated like-sign 
component originating from ${\rm B}$ meson decays, the ``like-sign" method to 
determine the background is inappropriate. 
We discuss strategies to deal with the correlations in the background.
By taking advantage of the ${\rm B}$ meson oscillation mechanism and of 
the particular features of ${\rm B}$ meson decays a new method 
to measure the ${\rm b}\bar{\rm b}$ production cross-section is proposed. 
\end{abstract}

\vspace{1.cm}

\pacs{PACS :}

\newpage

%%%%%%%%%%%%%%%%%%%%%%%%%%%%%%%%%%%%%%%%%%%%%%%%%%%%%%%%%%%%%%%%%%%%%%%%%%%
\section{Introduction}

Heavy quarkonia states provide, via their leptonic decays, an essential tool
to probe the earliest stage of heavy ion collisions at ultra-relativistic 
energies and are considered a crucial signature for demonstrating the 
existence of the quark gluon plasma (QGP) and for diagnosing its 
properties~\cite{matsuiPLB86,satzREP00,ramona3,ramona}.
The anomalous suppression of the $J/\psi$ meson, observed in ${\rm Pb+Pb}$
reactions at CERN/SPS, is indeed considered as an indication for the creation 
of a deconfined medium~\cite{na50}.
However, the production mechanism for both open and hidden charm at SPS is 
a subject of intensive current debate.
Thus the standard pQCD-based framework~\cite{matsuiPLB86,satzREP00} for 
charmonium production and interaction with the deconfined medium is nowadays 
confronted with new approaches such as 
the statistical hadronization model~\cite{pbmPLB00}
and the kinetic recombination model~\cite{thews01}.
These statistical models differ from the standard approach in their 
predictions for higher beam energies: they imply a significant $J/\psi$ 
enhancement rather than suppression.
In both approaches rescattering of heavy quarks in a deconfined medium is
essential. 
Furthermore, a unique signature of the new mechanisms is the peculiar 
centrality dependence~\cite{pbmNPA01}, a scaling of the quarkonia yield
with the square of the open charm (or bottom) 
yield~\cite{pbmPLB00,thews01,pbmNPA01} and thermal ratios for relative yields
of quarkonia states~\cite{pbmPLB00}.
It then becomes mandatory to measure, in addition to the quarkonia yields,
also the yields for open charm and open beauty.
The study of high mass dileptons is therefore one of the major physics goals
of RHIC and the LHC heavy ion program.
This new regime of beam energies brings however the challenge of 
extracting, for the first time, the quarkonia signals in the presence of a 
significant and highly complex combinatorial background which arises mostly 
from the semi-leptonic decay of open charm and open bottom 
(up to 200 ${\rm c}\bar{\rm c}$ and 6 ${\rm b}\bar{\rm b}$ pairs are 
expected to be produced per central ${\rm Pb+Pb}$ collision at LHC energy).
Consequently, high mass dilepton spectra exhibit peculiar features and 
the well known techniques for background subtraction which work successfully 
for low mass resonances at low beam energies, cannot be applied in a 
straightforward way.  

Based on simulations performed for ${\rm Pb+Pb}$ collisions at LHC energy 
we discuss the problems related to the subtraction of the combinatorial 
background to the dilepton spectra at high invariant mass.
We demonstrate that the peculiar characteristics of the combinatorial 
background at high invariant mass provides a means to measure, albeit
indirectly, the number of ${\rm b}\bar{\rm b}$ pairs produced in the reaction.

\section{Subtracting the combinatorial background}

In the multiparticle environment characteristic of heavy ion reactions
the possibility to observe a resonance relies on the fact that 
the reconstructed invariant mass of its decay products appears as a 
narrow peak signal superimposed on a broad background.
Depending on the underlying physics and on the event multiplicity, 
the background originates from uncorrelated particles and/or from correlated 
particles i.e. of common origin.
In principle the signal can be extracted by fitting the 
signal+background distribution with appropriate functions chosen 
in order to provide a good description of the overall spectrum.
This technique, however, does not work if signal and background have a 
similar shape.
This could result either i) from a bump in the background due to physical 
reasons, detector acceptance, or analysis cuts, or ii) from a broad signal 
due to a large intrinsic resonance width and/or a smearing of this width 
by the apparatus resolution.
In these cases it becomes difficult or even impossible to disentangle the 
signal and background components by fitting their sum.
Obviously, the situation gets even worse when the resonant signal is small 
and/or when the shape of the background is unknown.
These difficulties become even more obvious when extracting a continuum like 
the Drell-Yan, the thermal radiation, or the open charm/bottom dilepton signal.
The only way to overcome this problem is to estimate independently the 
background distribution and to subtract it from the signal+background spectrum.
Such a technique is nowadays intensively used in heavy ion collisions at 
high beam energies.

Considering an unlike-sign two particle invariant mass spectrum, there are 
mainly two ways to determine and subtract the combinatorial background:
\begin{itemize}
\item{Like-sign pair technique: the uncorrelated background in the 
unlike-sign pair sample is estimated by the number of like-sign pairs 
within each event.
The signal {\rm S} of the number of unlike-sign pairs ${\rm N}^{+-}$
is then given by 
${\rm S}~=~{\rm N}^{+-}~-~2 {\rm R}\sqrt{{\rm N}^{++}{\rm N}^{--}}$
where the factor {\rm R} accounts for a possible asymmetry in the production 
of positively and negatively charged particle and/or an asymmetry due to a 
detector trigger/acceptance bias relative to the particle 
charge. 
This technique has the disadvantage that the statistics in the background
spectrum is limited to the number of available events.
On the other hand, since the number of unlike-sign pairs and 
like-sign pairs are calculated within the same event, the normalization
of the determined background to the signal+background spectrum is 
straightforward, provided that one knows the ${\rm R}$ factor with good 
accuracy.}
\item{Event-mixing technique: the uncorrelated background in the 
unlike-sign pair sample is given by an unlike-sign pair spectrum for which 
the two particles of a pair are taken from two different events.
This offers a better statistical precision than the previous method
since one can mix each event with several (or even many) other events.
The mixed events have to be identical in centrality. 
In presence of flow they have to be rotated into a common reaction plane.
After normalization of the signal+background spectrum and the event-mixing 
spectrum to the respective number of events an additional normalization factor
of 2 has to be applied to the event-mixing spectrum. 
Note that with real data, the normalization factor is not straightforward 
because, due to fluctuations and detector resolution, the centrality of 
the mixed events is never strictly identical.
Therefore the normalization of the estimated background to the 
signal+background is usually done from the integration of the two spectra in 
a region where the correlated signal is assumed to be negligible.
Note also that, in order to be applied to experimental data, the
event-mixing technique requires that the two track resolution of the 
detector be taken into account.}
\end{itemize}
Each method has its own advantages and drawbacks depending on the particle 
environment and on the nature of the signal to be 
extracted~\cite{lhote94,marek,nick}.
In particular, the event-mixing technique can lead to incorrect results in 
the case of large signal-to-background. 
This can be intuitively understood by considering the extreme case of a 
sample of events where each event consists of two correlated particles only,
such as a ${\rm c}\bar{\rm c}$ pair.
In this case there would be no real background to the signal 
but the event-mixing technique would produce a fake background.
On the other hand it is obvious that the like-sign method, which relies on the 
fact that a like-sign pair is always uncorrelated, will fail if the events
contain like-sign correlated particle pairs.

In the following we shall show that, because the decay products of bottom
decay contain like-sign correlated lepton pairs, the applicability
of the like-sign technique for background subtraction in dilepton
physics at the collider energies, such as currently available
at RHIC and to become available at the LHC, is questionable.
By making use of the differences between the background estimated with the 
like-sign and the event-mixing techniques, we propose a new method to 
measure indirectly the number of ${\rm b}\bar{\rm b}$ pairs from the dilepton 
spectra.
As illustrative examples, we consider the two dilepton channels 
(${\rm e}^+{\rm e}^-$, $\mu^+\mu^-$) in the acceptance of the 
ALICE detector~\cite{alicetp} for central ${\rm Pb+Pb}$ collisions.
Technical details about the simulation are given in appendix A.

\section{Discussion}

Figure~\ref{imass1} shows the invariant mass (${\rm M}$) distribution of 
unlike-sign lepton pairs in the ALICE detector acceptance.
With a transverse momentum threshold of $2~{\rm GeV/c}$ on each lepton, 
the dileptons from ${\rm B}$ meson decays are the dominant component of the 
combinatorial background all over the invariant mass region except 
for $3 \lesssim {\rm M} \lesssim 5~{\rm GeV/c^2}$ where the component from
${\rm D}$ meson decays is significant.
Particularly interesting is the difference in shape between the two invariant 
mass distributions: the distribution from ${\rm D}$ meson decays 
exhibits one single maximum at intermediate ${\rm M}$ and a shoulder at 
low ${\rm M}$ while that from ${\rm B}$ meson decays shows two bumps at 
low and intermediate ${\rm M}$.
The bump at intermediate ${\rm M}$ results mostly from a combination
of two primary leptons while the bump at low ${\rm M}$ is a result of a 
combination of a primary and a secondary lepton produced by the same ${\rm B}$ 
meson in the so-called ${\rm B}$-chain channel\footnote{A primary lepton 
${\rm l}_1$ is a lepton directly produced by the meson ${\rm Q}$ in the first 
decay generation: ${\rm Q} \rightarrow {\rm l}_1$ + anything.
A secondary lepton ${\rm l}_2$ is a lepton produced in the second decay
generation: 
${\rm Q} \rightarrow {\rm X}$ + anything,
${\rm X} \rightarrow {\rm l}_2$ + anything.}.
This channel represents a sizeable fraction of dileptons 
because the first generation of decay products of a ${\rm B}$ meson generally 
contains a ${\rm D}$ meson whose semi-leptonic decay branching ratio is 
rather large.
On the other hand, this channel is almost inexistent for primary ${\rm D}$ 
mesons because the secondary leptons from charm decay mostly originate from 
$\pi$ and ${\rm K}$ meson decay.
Most of these secondary leptons which have very low momenta are removed by 
the transverse momentum threshold.
Consequently, the dilepton distribution from ${\rm D}$ meson decay
is dominated, to a large extent, by pairs of primary leptons.

\begin{figure}[hhh]
\includegraphics{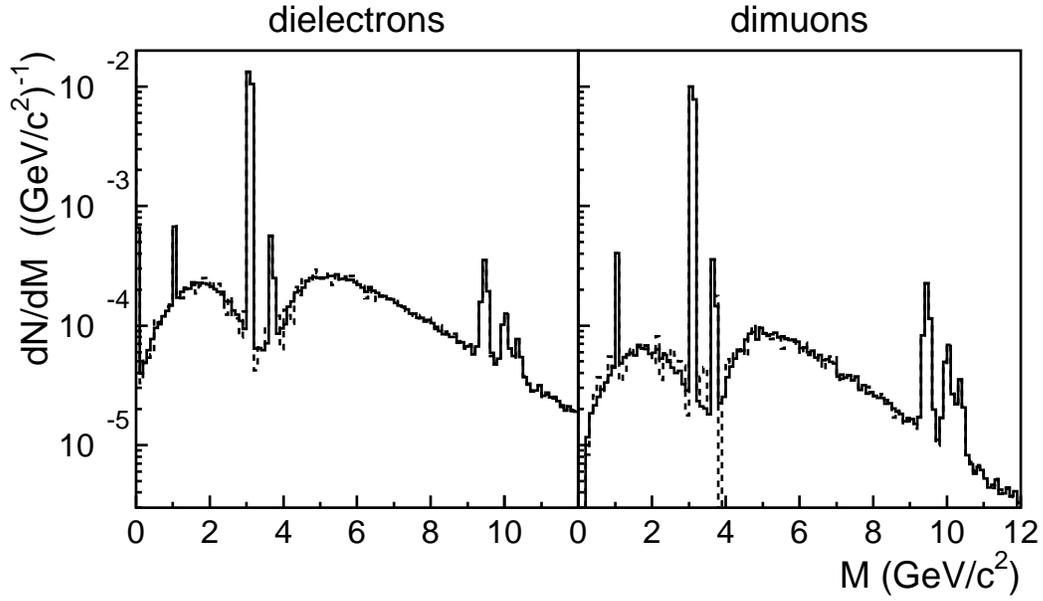}
\vspace*{16.cm}
\caption{Invariant mass distributions of unlike-sign correlated 
dileptons (solid histograms) for central ${\rm Pb+Pb}$ reactions at LHC energy.
The events are filtered through the ALICE detector acceptance cuts as 
described in the text.
The dashed histogram is obtained by the subtraction of the background,
determined by the event-mixing technique, from the total distribution.
For more details, see text.}
\label{imass2_em}
\end{figure}

\begin{figure}[hhh]
\includegraphics{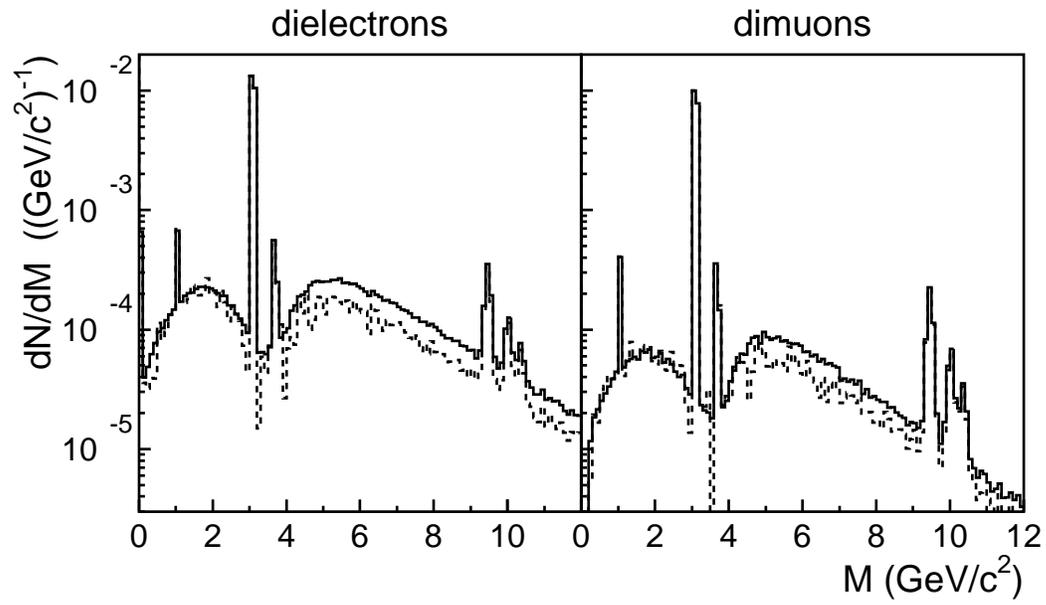}
\vspace*{16.cm}
\caption{Same as Fig.~\ref{imass2_em} but using the like-sign technique
instead of the event-mixing technique for the determination of the background.}
\label{imass2_ls}
\end{figure}

The background to the unlike-sign correlated dilepton signal is
estimated by means of the two previously described techniques.
The like-sign spectrum is normalized with the ${\rm R}$ factor equal to 1
since no charged particle asymmetry is present in our simulated events
(more details about the ${\rm R}$ factor are given in appendix B)
Concerning the event-mixing spectrum we use the standard normalization 
factor of 2 mentioned previously since all simulated events can be considered
as strictly identical in terms of centrality.
The resulting background distributions are shown by the dashed and the dotted 
histograms in Fig.~\ref{imass1}.
Clearly the two techniques do not lead to the same 
result: for large ${\rm M}$, the background from the like-sign method 
systematically underestimates that from the event-mixing method.
This can be better observed by inspecting Fig.~\ref{imass2_em} 
and~\ref{imass2_ls} 
where the background-subtracted spectrum is compared to the sum of all 
unlike-sign correlated signals.
Figure~\ref{imass2_em} shows that the event-mixing technique works as 
expected i.e. the background-subtracted spectrum is identical to the 
sum of the unlike-sign correlated signals (the difference between the
solid and the dashed histograms is hardly visible in Fig.~\ref{imass2_em}).
On the other hand, Fig.~\ref{imass2_ls} shows that the background-subtracted 
spectrum with the like-sign technique seems to describe well the unlike-sign 
correlated signal at low ${\rm M}$ but underestimates this signal by about 
$25\%$ for ${\rm M} \gtrsim 3~{\rm GeV/c}^{\rm 2}$.
%%%%%%%%%%%%%%%%%%%%%%%%%%%%%%%%%%%%%%%%%%%%%%%%%%%%%%%

%%%%%%%%%%%%%%%%%%%%%%%%%%%%%%%%%%%%%%%%%%%%%%%%%%%%%%%
Two effects contribute to the failure of the like-sign technique. 
Both of them are related to the particularity of the ${\rm B}$ mesons in a 
sense that a ${\rm B}\bar{\rm B}$ pair produces not only unlike-sign 
correlated lepton pairs, but also like-sign correlated lepton pairs:
\begin{itemize}
\item{The first decay generation of ${\rm B}$ mesons contains 
$\sim $~10$\%$ of primary leptons and a large fraction of ${\rm D}$ 
mesons which decay semi-leptonically with a branching ratio of $\sim$~12$\%$.
Therefore a ${\rm B}\bar{\rm B}$ pair is a source of like-sign correlated 
pairs.
For example, in the following decay chain:\\
\hspace*{2.cm}${\rm B}^+$ $\rightarrow$ 
$\bar{\rm D}^0$ ${\rm e}^+$ $\nu_{\rm e}$,
$\bar{\rm D}^0$ $\rightarrow$ ${\rm e}^-$ anything\\
\hspace*{2.cm}${\rm B}^-$ $\rightarrow$ ${\rm D}^0$ $\pi^-$,
${\rm D}^0$ $\rightarrow$ ${\rm e}^+$ anything\\
the ${\rm B}^+{\rm B}^-$ pair produces a correlated ${\rm e}^+{\rm e}^+$ 
pair in addition to the ${\rm e}^+{\rm e}^-$ pair.}
\item{Like the ${\rm K}^0\bar{\rm K}^0$ system, 
the 2 neutral ${\rm B}^0\bar{\rm B}^0$ meson systems 
${\rm B}^0_{\rm d}\bar{\rm B}^0_{\rm d}$ and 
${\rm B}^0_{\rm s}\bar{\rm B}^0_{\rm s}$ 
undergo the phenomenon of particle-antiparticle mixing (or oscillation).
This effect is quantified by the so-called mixing parameter
$\chi_{\rm d}$ ($\chi_{\rm s}$) which corresponds to the time-integrated
probability that a produced ${\rm B}^0_{\rm d}$ (${\rm B}^0_{\rm s}$) decays 
as a $\bar{\rm B}^0_{\rm d}$ ($\bar{\rm B}^0_{\rm s}$) and vice versa.
The mixing parameters are predicted by the Standard Model and have been 
measured experimentally~\cite{booklet}.
They are estimated to $\chi_{\rm d}$ $=$~0.17 and $\chi_{\rm s}\ge$~0.49.
Therefore, a ${\rm B}^0_{\rm d}\bar{\rm B}^0_{\rm d}$ 
(${\rm B}^0_{\rm s}\bar{\rm B}^0_{\rm s}$)
pair produces, in the primary dilepton channel,
$\sim$~70$\%$ ($\sim$~50$\%$) of unlike-sign correlated lepton pairs and 
$\sim$~30$\%$ ($\sim$~50$\%$) of like-sign correlated lepton pairs.}
\end{itemize}
Due to these effects the unlike-sign dilepton spectrum does not contain the 
full correlated signal because a part of this correlated signal is made of 
like-sign lepton pairs.
Consequently, the like-sign subtraction removes from the unlike-sign 
dilepton spectrum not only the uncorrelated component but also a fraction 
of the correlated signal.
Note that, in the case presented in Fig.~\ref{imass2_ls}, the use of the 
like-sign technique to subtract the background to the $\Upsilon$ signal 
would bias only weakly the result because the yield in the resonances 
peak is much larger than the yield in the background below these peaks.
On the contrary, this bias would become important in the case of a 
strong $\Upsilon$ suppression which would manifest itself by an almost 
vanishing peak.
It is obvious that it would become even more important for continuum physics.

\begin{figure}[hhh]
\includegraphics{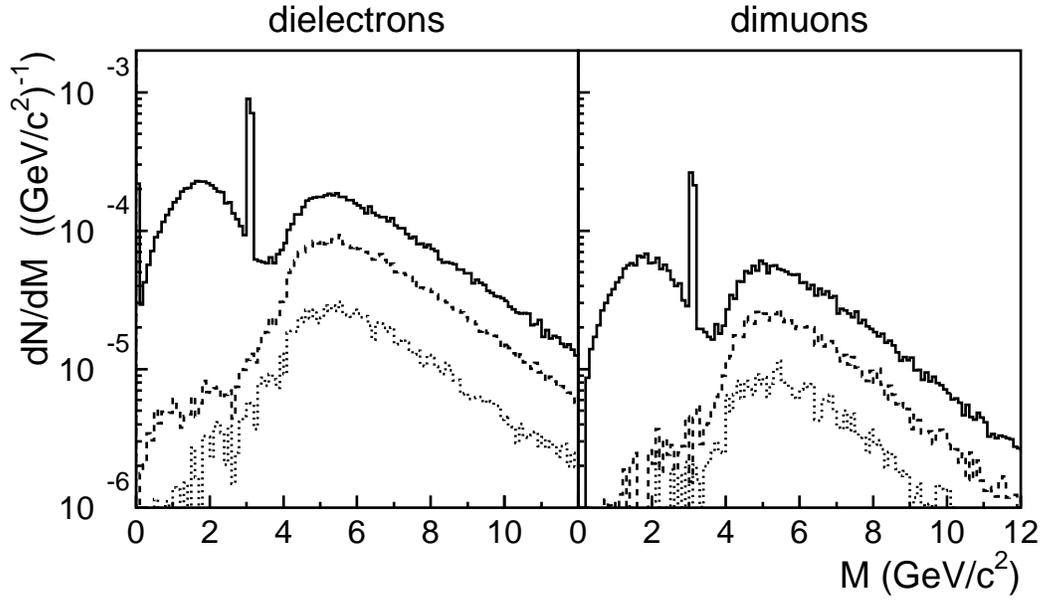}
\vspace*{17.cm}
\caption{Invariant mass distributions of correlated dileptons from 
bottom decay for central ${\rm Pb+Pb}$ reactions at LHC energy. 
The events are filtered through the ALICE detector acceptance cuts as 
described in the text.
The unlike-sign and like-sign components are shown by the solid and 
dashed histograms, respectively.
The dotted histogram represents the like-sign component which results from 
${\rm B}^0\bar{\rm B}^0$ mixing.}
\label{imass4}
\end{figure}

Figure~\ref{imass4} shows the like-sign and the unlike-sign correlated 
components.
For ${\rm M}~\lesssim~3~{\rm GeV/c}^2$, the correlated ${\rm B}$ meson decay 
distribution almost exclusively consists of unlike-sign pairs.
For ${\rm M}~\gtrsim~3~{\rm GeV/c}^2$, the like-sign correlated component 
reaches about $45\%$ of the unlike-sign correlated component.
The like-sign correlated component arising from ${\rm B}^0$ mixing 
amounts to about $30\%$ of the total like-sign correlated distribution.

\begin{figure}[hhh]
\includegraphics{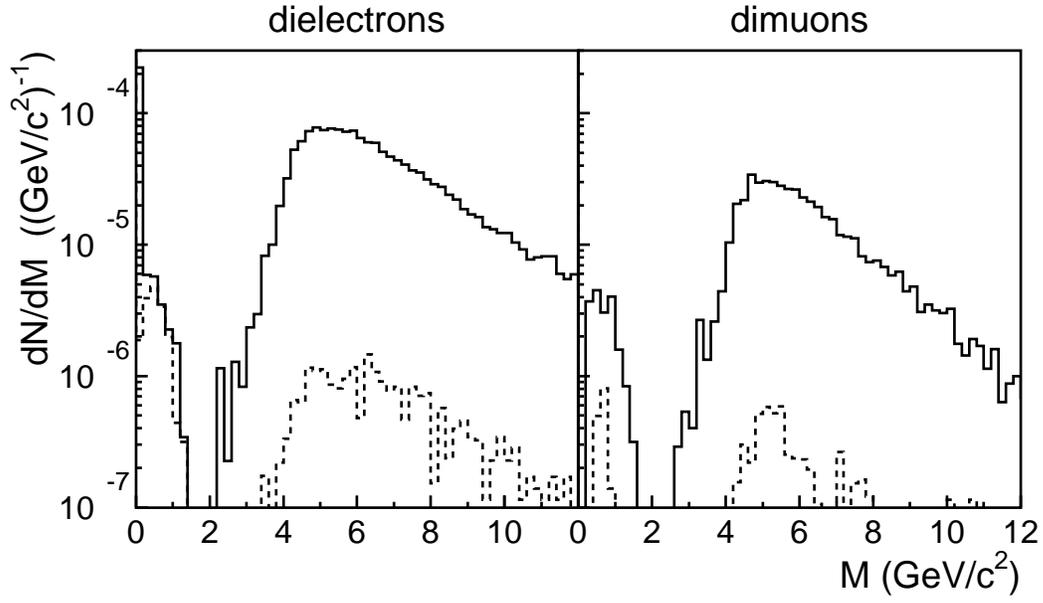}
\vspace*{17.cm}
\caption{Invariant mass distributions of correlated dileptons from 
charm decay for central ${\rm Pb+Pb}$ reactions at LHC energy.
The events are filtered through the ALICE detector acceptance cuts as 
described in the text.
The unlike-sign and like-sign components are shown by the solid and 
dashed histograms, respectively.}
\label{imass5}
\end{figure}

Since ${\rm D}$ mesons do not oscillate (no evidence for ${\rm D}$ meson 
oscillation has been observed so far~\cite{booklet}), 
a ${\rm D}\bar{\rm D}$ pair cannot produce a like-sign correlated lepton 
pair in the primary decay generation.
Nevertheless, ${\rm D}$ mesons can generate like-sign correlated 
lepton pairs from their decay chain.
It can be seen from Fig.~\ref{imass5} that the corresponding size of the 
effect
is very weak in both the dielectron channel and the dimuon channel.
As already mentioned, this is due to the fact that most of the secondary
leptons from charm decay are removed by the transverse momentum threshold 
of 2~{\rm GeV/c}.

\begin{figure}[hhh]
\includegraphics{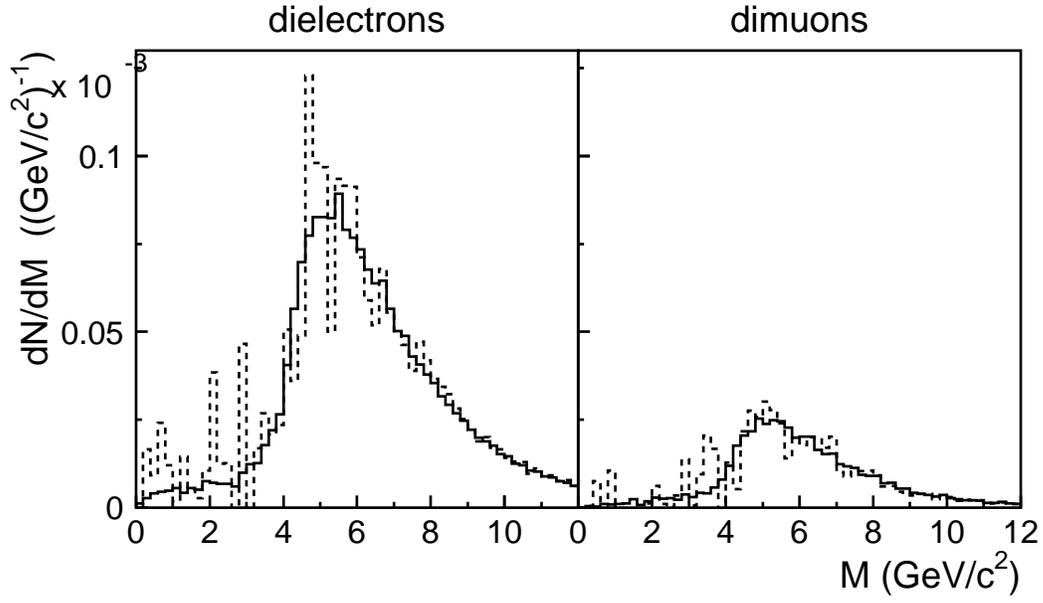}
\vspace*{17.cm}
\caption{Invariant mass distributions of dileptons for central 
${\rm Pb+Pb}$ reactions at LHC energy.
The events are filtered through the ALICE detector acceptance cuts as 
described in the text.
Solid histogram: like-sign correlated dileptons from ${\rm B}$ meson decays.
Dashed histogram: distributions obtained after subtraction of the background 
estimated with the event-mixing technique from the one estimated by the 
like-sign technique.}
\label{imass3}
\end{figure}

As shown above, the event-mixing background does not contain any correlated 
signal whereas the like-sign background contains, in addition to the 
uncorrelated signal, the like-sign correlated lepton pairs from bottom decay.
Therefore, subtracting the first background from the second one should 
get access to the like-sign correlated component.
Figure~\ref{imass3} shows that this is indeed the case in the present 
simulations: 
when subtracting the event-mixing spectrum from the like-sign spectrum, 
one obtains a non-zero distribution which corresponds precisely to the 
like-sign correlated component from bottom decay.
The latter is directly connected to the full ${\rm b}\bar{\rm b}$
cross-section since any change in the ${\rm b}\bar{\rm b}$ cross-section 
will translate into a proportional change in the like-sign correlated
dilepton yield.
The signal obtained from the subtraction of the two background distributions
can therefore be considered as a reliable measurement of the number of 
${\rm b}\bar{\rm b}$ pairs.
Indeed, neither a resonance nor the thermal radiation nor the Drell-Yan 
mechanism can provide like-sign correlated dileptons and, 
as shown in Fig.~\ref{imass5}, the amount of like-sign correlated lepton pairs
from charm decay is negligible when applying a $p_{\rm t}$ threshold
of 2~{\rm GeV/c} on both leptons. 

Note that one could also get a model-independent estimate of the number of 
${\rm b}\bar{\rm b}$ pairs from the integration of the background-subtracted 
unlike-sign spectrum at high invariant mass where the dileptons from bottom 
decay dominate (Fig.~\ref{imass1}).   
However, this could lead to a non-pure bottom sample since the 
integrated signal would contain all kind of unlike-sign correlated pairs.
In particular, it would contain the Drell-Yan signal whose yield could 
become significantly large relative to the sum of correlated charm and bottom 
yields in the case of strong energy loss effects on heavy 
quarks~\cite{kamPLB98}.
It would also contain the unlike-sign correlated pairs from charm whose 
thermal production cross-section could be enhanced for 
a QGP with a relatively high temperature~\cite{shuryak92,geiger93},
as well as lepton pairs from a thermalized hot QGP.

We point out that whereas in ${\rm e}^+{\rm e}^-$ reactions a 
good understanding of open beauty hadron production, sample 
composition and decay has been recently achieved~\cite{booklet}, in 
nucleus-nucleus reactions none of the open beauty hadrons has ever been 
measured so far (this holds even for open charm hadrons).
For the present investigations we have assumed that B meson production,
mixing, and decay can be extrapolated from nucleon-nucleon collisions to
nucleus-nucleus collisions by multiplication with the number of binary
collisions. The reality may be far from this, and it is indeed a major
goal of the ultra-relativistic heavy ion program to study the
differences. However, for the present studies which concern mostly the
signal/background we believe that the current investigation addresses
the crucial and important new points which can be studied in the heavy
ion environment.

Note, finally, that the proposed method provides a global estimate of the 
like-sign signal from bottom including the signal from ${\rm B}^0$ mixing 
and from the ${\rm B}$ meson decay chain.
Further detailed investigations should reveal whether the two contributions 
might be identified separately by means of further constraints on the 
kinematical characteristics of the lepton pairs.

\section{Conclusions}

We have presented some features of high mass dilepton spectra in heavy 
ion collisions at collider energies with special emphasis on the determination 
of the continuum background at high invariant masses.
Our investigations show that the combinatorial background contains a large 
amount of dileptons from bottom decay.
Because bottom decay is a source of like-sign correlated lepton pairs, the 
result of the background determination using the like-sign technique is 
inappropriate.
The event-mixing method is not affected by this effect 
and gives a reliable estimate of the combinatorial background.
We have demonstrated that, by subtracting the event-mixing distribution
from the like-sign distribution, one obtains a precise estimate of 
the yield of like-sign correlated lepton pairs from bottom decay.
These results are relevant for analysis of data from RHIC and the LHC.

\section*{Acknowledgements}

We thank D.~Mi\'skowiec for careful reading of the manuscript and for
useful comments.

\section*{Appendix A: simulation environment}

Our simulation is similar to that presented in~\cite{lin99}.
It consists of the following steps:
\begin{itemize}
\item{Open charm and open bottom in central ${\rm Pb+Pb}$ reactions are 
computed with PYTHIA~5.7~\cite{sjo94} assuming that a ${\rm Pb+Pb}$ 
collision is a superposition of a certain number of {\rm p+p} reactions.
We use the GRVH0 parton distribution function to normalize 
the production cross-sections in {\rm p+p} reactions at 
$\sqrt{\rm s}~=~5.5~{\rm TeV}$ to 6.7~${\rm mb}$ 
and 0.2~${\rm mb}$ for charm and bottom respectively~\cite{gavaiINT95}.
These cross-sections are then extrapolated to ${\rm Pb+Pb}$ collisions
at impact parameter $b~=~0~{\rm fm}$ with the nuclear overlap function  
${\rm T}_{\rm PbPb}(0)~=~30.4~{\rm mb}^{-1}$~\cite{ramona2}.
It results into 205 ${\rm c}\bar{\rm c}$ and 6 ${\rm b}\bar{\rm b}$
per central ${\rm Pb+Pb}$ event.}
\item{The heavy quarks hadronize into heavy mesons through 
a Peterson fragmentation function~\cite{peterson}. 
The ${\rm c}$($\bar{\rm c}$) quarks are assumed to fragment into 
${\rm D}^+$(${\rm D}^-$), ${\rm D}^0$($\bar{\rm D}^0$), and 
${\rm D}^+_{\rm s}$(${\rm D}^-_{\rm s}$) mesons.
For the ${\rm b}$($\bar{\rm b}$) quarks, we consider 
${\rm B}^+$(${\rm B}^-$), ${\rm B}^0$($\bar{\rm B}^0$),
${\rm B}^0_{\rm s}$($\bar{\rm B}^0_{\rm s}$),
and $\Lambda^0_{\rm b}$($\bar{\Lambda^0_{\rm b}}$).}
\item{Heavy mesons decay according to JETSET~6.4~\cite{sjo94} 
with free decay branching ratios.}
\item{Resonance events including $\phi$, $J/\psi$, 
$\psi^{\prime}$, $\Upsilon$, $\Upsilon^{\prime}$ and $\Upsilon^{\prime\prime}$
are generated separately with yields and spectra taken 
from~\cite{trd99,spectro96}. 
The resonances decay according to JETSET with fixed branching ratio.}
\item{The ALICE detector response is modeled in a simple way 
by\footnote{We do not intend to reproduce detailed features of the ALICE 
detector response but use representative numbers for illustrative purposes.}:
\begin{itemize}
\item{geometrical conditions: $45^\circ~<~\Theta_{\rm e}~<~135^\circ$ and
$2^\circ~<~\Theta_{\mu}~<~9^\circ$ where $\Theta_{\rm e}$ and 
$\Theta_{\mu}$ denote the polar angle of electrons and muons, respectively;}
\item{lower and upper transverse momentum thresholds: 
$2~<~p_{\rm t}~<~10~{\rm GeV/c}$ for both electrons and muons;}
\item{conditions on vertex: $v_{\rm t}^{\rm e}~<~3~{\rm cm}$ and 
$v_{\rm z}^\mu~<~100~{\rm cm}$ where $v_{\rm t}^{\rm e}$ and 
$v_{\rm z}^\mu$ are the electron transverse distance and muon 
longitudinal distance between the interaction point and the track vertex.
The second vertex cut simulates the effect of the front absorber of the muon
spectrometer;}
\item{transverse momentum resolution:
$\Delta p_{\rm t}/p_{\rm t}$ increases linearly from 0.6$\%$ at 
$p_{\rm t}$=1~GeV/c to 1.4$\%$ at 
$p_{\rm t}$=10~GeV/c for both electrons and muons.}
\end{itemize}}
\item{The leptonic component of central ${\rm Pb+Pb}$ events is 
constructed by means of a cocktail of the generated leptons from open charm, 
open bottom and resonances.
Each lepton is given a weight which includes the production 
cross-section of its source folded with the actual number of simulated  
sources per event and the corresponding decay branching ratio.
The weight and the momentum components of the leptons are registered keeping 
the information on the direct parent and the grandparent of the decay product. 
This allows to trace back the composition of final spectra.}
\end{itemize}
Leptons from other sources are neglected.
Not taken into account in the simulation are:
shadowing of the structure functions, energy loss of the partons in the medium,
rescattering of heavy mesons, suppression/enhancement of the resonance yields.
The ALICE detector filter assumes perfect particle identification and 
efficiency. 

\section*{Appendix B: origins of charge asymmetry and the ${\rm R}$ factor}

The presence of an asymmetry in the production of positive and negative 
charged background leptons leads to an enhancement of the like-sign dileptons
relative to the unlike-sign dileptons.
Therefore, the correct normalization of the like-sign spectrum relies on a 
precise determination of the ${\rm R}$ factor.
We discuss here the origins of possible charge asymmetries in nucleus-nucleus
reactions, in which conditions this leads to a ${\rm R}$ factor different from
unity and how to estimate ${\rm R}$ in this case.

\subsection*{Charge asymmetry at the production level}

At least three physical effects at the particle production level can lead
to a charge asymmetry in the final stage of nucleus-nucleus reactions:
\begin{itemize}
\item{charge conservation implies an excess of positively charged hadrons 
in the final state because the colliding ions are positively charged;}
\item{isospin conservation implies a ratio $\pi^-/\pi^+ >1$ because the 
ion ${\rm N}/{\rm Z}$ ratio is larger than 1;}
\item{associate Kaon production results in an excess of ${\rm K}^+$ relative 
to ${\rm K}^-$.}
\end{itemize}
It is obvious that these asymmetries propagate to the decayed leptons.
However in central nucleus-nucleus reactions the multiplicity of charged 
particles produced in the interaction is so large that any initial charge 
asymmetry is smeared-out~\cite{na503}.
In fact, it can be demonstrated that ${\rm R}=1$ exactly if the particle 
multiplicities are Poisonnian~\cite{bellaiche}.
This has also been shown by means of Monte-Carlo simulations for central
${\rm Pb+Pb}$ reactions at SPS~\cite{na503}.
The charge asymmetry tends to vanish with increasing beam energy as evidenced 
by the measured $\pi^-/\pi^+$ ratio which goes, in central ${\rm Au+Au}$ 
reactions, from 2.15$\pm$0.30 at GSI/SIS~\cite{fopi}
to 1.00$\pm$0.02 at RHIC~\cite{phobos}.

On the contrary, the previous statement is not valid anymore in low 
multiplicity events like non-central nucleus-nucleus reactions. 
In this case ${\rm R}$ is larger than unity.
Its actual value can be precisely determined by means of 
simulations~\cite{na503}.
Even in this case ${\rm R}$ differs only slightly from unity.
Indeed in peripheral (${\rm b}\sim 13~{\rm fm}$) ${\rm Pb+Pb}$ collisions 
at SPS, the ${\rm R}$ factor for dimuons in the acceptance of the 
NA50 spectrometer is found to be 1.075~\cite{na503}.
It should be even closer to unity at higher beam energy thanks to the 
larger particle multiplicity.

\subsection*{Charge asymmetry due to a detector acceptance/trigger bias}

A charge asymmetry in the lepton sample can also be the consequence of a
possible detector bias relative to the lepton charge.
The detector bias caused by a different acceptance for positively and 
negatively charged particle is usually taken into account by averaging the 
data collected with the two opposite settings of the magnetic 
field~\cite{bellaiche}.
Similarly a charge asymmetry could result from a different probability 
for the trigger system to accept like-sign and unlike-sign events.
Such an effect can be estimated by means of simulations as discussed 
in~\cite{helios}.

\newpage

\begin{figure}[hhh]
\includegraphics{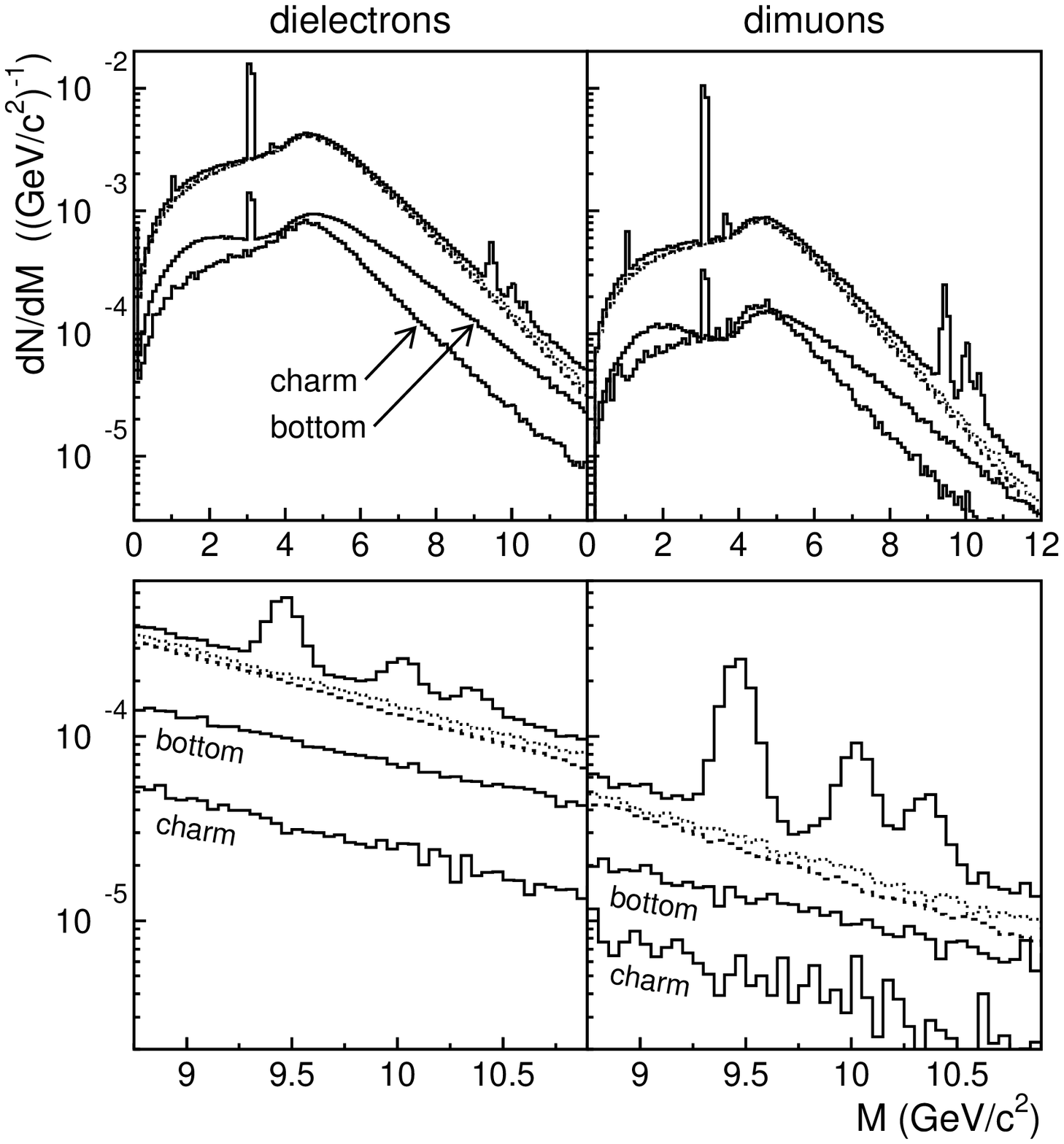}
\vspace*{16.cm}
\caption{Invariant mass distributions of unlike-sign dielectrons (left)
and unlike-sign dimuons (right) for central ${\rm Pb+Pb}$ reactions 
at LHC energy.
The events are filtered through the ALICE detector acceptance cuts as 
described in the text.
The top (solid) histogram on each figure comprises the sum of all unlike-sign 
lepton pairs.
The dashed and dotted histograms show the background determined by the 
event-mixing technique and with the like-sign technique, respectively.
The histograms labeled charm and bottom correspond to the component where 
both leptons result from ${\rm D}$ and ${\rm B}$ meson decay, respectively.
The two lower panels are zooms of the upper panels in the invariant mass
region of the $\Upsilon$ family.}
\label{imass1}
\end{figure}

\newpage

\newpage

\newpage

\newpage

\newpage

\end{document}